\documentstyle[aps,prl,epsf,floats]{revtex}
\begin{document} 
\input{epsf}
\input psfig.sty
\title{Self-organized criticality in the hysteresis of the Sherrington - 
Kirkpatrick model}
\author{Ferenc P\'azm\'andi$^{1,2,3}$, Gergely Zar\'and$^{1,3}$, and 
Gergely T. Zim\'anyi$^{1}$}
\address{$^1$Physics Department, University of California, Davis, CA 95616 \\
$^2$Physics Department,  KLTE, H-4010 Debrecen P.O. Box 5.Debrecen, Hungary \\
$^3$Research Group of the Hungarian Academy of Sciences, Institute of Physics, 
TU Budapest, H-1521 Hungary}

\twocolumn[\hsize\textwidth\columnwidth\hsize\csname@twocolumnfalse\endcsname
\date{\today}
\maketitle
\begin{abstract}
We study hysteretic phenomena in random ferromagnets. We argue that the 
angle dependent magnetostatic (dipolar) terms introduce
frustration and long range interactions in these systems. This makes it
plausible that the Sherrington - Kirkpatrick model may be able to
capture some of the relevant physics of these systems.
We use scaling arguments, replica calculations and large scale numerical 
simulations to characterize the hysteresis of the zero temperature SK model.
By constructing the distribution functions of the avalanche sizes,
magnetization jumps and local fields, we conclude that the system exhibits 
{\it self-organized criticality} everywhere on the hysteresis loop.

\end{abstract}
\vskip 0.3cm]

Hysteresis in ferromagnetic systems is a century old physical
problem. Efficient phenomenologies have already
been developed~\cite{bertottibook}, 
but an accepted microscopic theory is yet to be constructed.
In soft magnets, where domain wall motion dominates the physics, considerable 
progress has been achieved recently~\cite{bertotti,nattermann,stanley}.
In hard magnets domain nucleation, domain wall motion and their interaction
are all important. Hence they are better described on a more microsopic level
as an assembly of strongly interacting spins or hysterons~\cite{pokrovsky}. 
Quantitative insight to such systems has been gained recently through
studying the random field Ising model (RFIM)~\cite{sethna}.

However a key aspect of the physics of real systems is missing from the RFIM:
it does not include the long range dipolar (or magnetostatic) interactions.
While these are negligible on atomic scales relative to the exchange term,
they can dominate the collective behaviour of granular systems.
This is so because the dipolar interaction is long ranged, so it involves
every spin in the volume of the grains, whereas the exchange coupling scales 
only with the number of spins on the surface of the grain.
These dipolar forces are important: they prevent the roughening of the domain 
walls~\cite{stanley} and determine the size of the domains~\cite{bertottibook}.
Crucially, the sign of these interactions changes with the angle. This
introduces {\it frustration} into the system, which is not represented in 
the RFIM. 

To capture the influence of frustration on hysteretic phenomena, 
we study the simplest system, containing long-range frustrated interactions, 
the Sherrington-Kirkpatrick (SK) model. 
Early numerical work demonstrated that this model exhibits hysteresis
~\cite{levin,bertotti2}. However, in spite of its obvious 
importance, we could not find analytic studies of the 
hysteresis loop of the SK model. In this Letter we use scaling arguments, 
replica calculations and large scale numerical simulations to characterize the 
hysteresis of the zero temperature SK model.
By constructing the distribution functions of the avalanche sizes,
magnetization jumps and local fields, we conclude that the system exhibits 
{\it self-organized criticality} everywhere on the hysteresis loop.

The SK model consists of $N$ Ising spins ($\sigma=\pm 1$) on a fully connected
lattice, described by the Hamiltonian
\begin{equation}
{\cal{H}}=-\frac{1}{2}\sum_{i\ne j=1}^N J_{ij}\sigma_i\sigma_j-h\sum_{i=1}^N 
\sigma_i,
\label{Ham}
\end{equation}
where $J_{ij}$ is a random Gaussian number of zero mean and variance
$1/N$. Throughout the paper we work at $T=0$.

\begin{figure}
\epsfxsize=8cm
\hskip0.5cm
\epsfbox{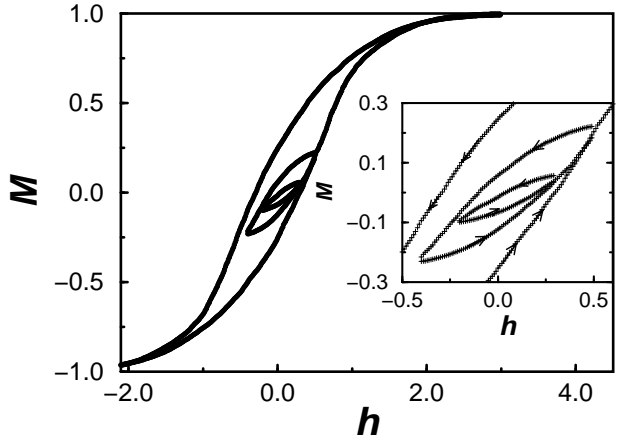}
\caption{\label{fig:hyst} The hysteresis loop of the SK model,
averaged over 100 disorder configurations ($N = 1600$).
Inset: multiple minor loops, exhibiting return point memory.}
\end{figure}

First we summarize our numerical results. We start from a fully polarized
state and change the external magnetic field $h$ adiabatically:
for a given field we let all spins align according to their local field before
varying $h$ again. During the avalanches  we use  sequential 
single spin flip updating to ensure the decrease of the total energy. 
The resulting hysteresis loop for the SK model is presented in 
Fig.~\ref{fig:hyst}. Finite size scaling analysis 
shows that the hysteretic trajectories are well-defined in the $N\to\infty$
limit, and the coercive field converges to a finite value. 

We also analyzed the minor hysteresis loops of the SK model (inset 
Fig.~\ref{fig:hyst}). Within numerical accuracy they return to the major 
loop at the point of departure, 
exhibiting {\it return point memory}. This feature is present in many
experimental systems, and it is also one of the criteria for the 
applicability of the Preisach phenomenology~\cite{bertottibook}.

Next we establish some of the basic energy scales
from elementary considerations. 
When spin $\sigma_j$ is flipped, the local field 
$h_i$ at another site changes by an amount proportional to
$2 J_{ij} \sim 2/N^{1/2}$. Thus the external field $h$ has to be changed
by an amount $dh \propto {1/N^{1/2}}$ to start a new avalanche.
Now let $S$ be the change in the {\it total} magnetization during
an avalanche, and $dm=S/N$ the jump of the magnetization
$m$ during the avalanche. The average $m(h)$ curve is continuous and thus 
its derivative $\langle dm/dh \rangle \propto \langle S \rangle/N^{1/2}$
is finite (Fig.~\ref{fig:hyst}), requiring:
$\langle S \rangle \propto N^{1/2}$.
This is possible only if the {\it scale of the distribution of avalanches}
is set by $N^{1/2}$. This is characteristic of systems {\it at criticality},
whereas off-criticality the scale is set by some control parameter
of the Hamiltonian. This leads to the central result of the paper: 
the SK model exhibits critical behaviour {\it everywhere} along its hysteresis 
loop. As this phenomenon is independent of the parameters of the Hamiltonian, 
it is a manifestation of {\it self-organized criticality}.  

\begin{figure}
\epsfxsize=8cm
\hskip0.5cm
\epsfbox{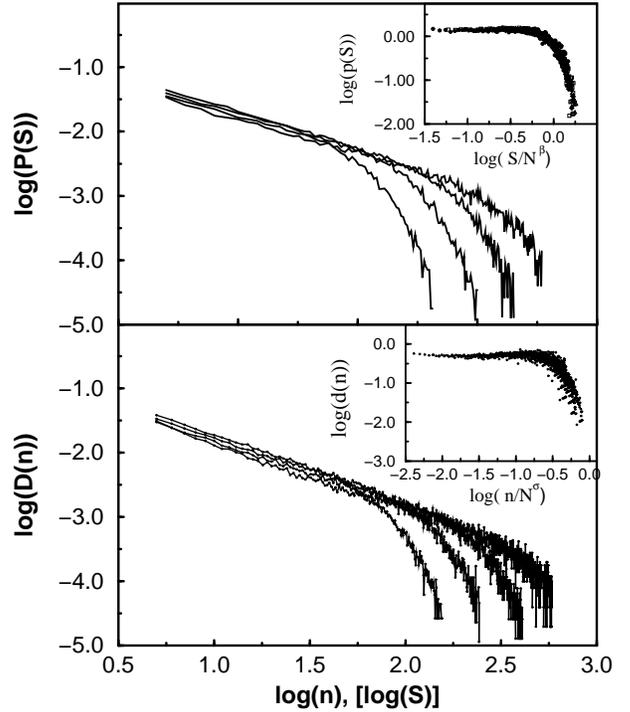}
\vskip0.01truecm
\caption{\label{fig:DnPs} Avalanche size and magnetization jump 
distributions ${\cal D}(n)$ and ${\cal P}(S)$ for system sizes 
$N = 400$, 800, 1600, and 3200. The inserts  show the collapsing scaling 
curves corresponding to Eq.~(\protect{\ref{eq:Ps}}) with $d(n/N^\sigma) = 
n \ln(N) {\cal D}(n)$ and $p(S/N^\beta) = 
S \ln(N) {\cal P}(S)$.}
\end{figure}

To elucidate this point, in Fig.~\ref{fig:DnPs} we show the distribution
functions of $S$, and the number of spin flips in an avalanche (its ``size''), 
$n$; ${\cal P}(S)$ and ${\cal D}(n)$, respectively, measured in the interval 
$m\in [-0.3,0.3]$ for various system sizes. Both distributions exhibit power 
law behavior and can be well described by the finite size scaling forms:
\begin{eqnarray}
{\cal D}(n) = (B/\ln N) \;n^{-\varrho} \;d(n/N^\sigma)\;,
\label{eq:Dn}\\
{\cal P}(S) = (A/\ln N) \;S^{-\tau} \;p(S/N^\beta)\;, 
\label{eq:Ps}
\end{eqnarray}
with $\tau, \varrho = 1 \pm 0.1$, $\sigma = 0.9 \pm 0.1$, and 
$\beta = 0.6 \pm 0.1$. The logarithmic prefactors were necessary 
to achieve satisfactory scaling collapse.
Since such terms are needed only to keep distributions
with an exponent $1$ normalized, this strongly suggests that
$\tau = \varrho = 1$ exactly. 
Unfortunately, because the cutoffs of the distributions 
${\cal P}(S)$ and ${\cal D}(n)$ scale with different powers of $N$
the attractive picture of a diffusive 
motion of the local fields due to the randomness in $J_{ij}$~\cite{bertotti} 
would lead to an infinite diffusion constant $D\propto \langle 
n\rangle/N^{1/2}$ and is thus inapplicable.

Adopting the $\tau = 1$ equality and combining it with 
$\langle S \rangle\sim N^{1/2}$ immediately yields the relation $\beta = 1/2$,
with logarithmic corrections, in good agreement with the above measured value.
Also, because the $J_{ij}$'s take negative values as well, spins of
{\it both} signs are destabilized in an avalanche. Therefore the number of
participating spins is only bounded from below by $S/2$, yielding the
exponent - bound $\sigma \ge \beta = 1/2$.
An upper bound for  
$\sigma$ can be  obtained from estimating the dissipated energy, 
$E_d$, during a finite but small sweep of the external-field 
$h_1\rightarrow h_2 = h_1 + \Delta h$: $E_d  = N m \Delta h\sim N$. 
Also, since  the average energy dissipation per spin is {\it at 
least} $2 dh \sim 1/N^{1/2}$,
$E_d$ can be estimated as $E_d>2 n_{total}/N^{1/2}$, where $n_{total}$
is the number of flips during all avalanches from $h_1$ to $h_2$.
But the number of avalanches during this sweep is proportional to
$\Delta h/dh \sim N^{1/2}$, i.e.  $n_{total}\sim N^{1/2} \langle n\rangle$. 
Combining all the above gives $N \sim E_d > \langle n\rangle\sim N^\sigma$ 
implying the upper bound $\sigma \le 1$, which is nearly saturated 
according to our numerics.

The above distributions imply that the {\it average} value of 
$\chi \equiv dm/dh$ is dominated by a few very large avalanches, whereas its 
{\it typical} value scales to zero as $\sim 1/N^{1/2}$, which we 
confirmed independently numerically. Therefore the hysteresis loop for a
specific disorder realization has a slope zero with unit probability,
interrupted by a few macroscopically large avalanches. This feature
is characteristic of the Barkhausen noise and establishes the frustrated spin
glasses as possible candidates to describe certain classes of
hysteretic magnets.

We also studied the correlations of consecutive avalanches. We measured the 
Hausdorf dimensions of the numerically determined hysteresis loop and
that of a sequence of independent avalanches, generated with the above 
distributions. Having found the two Hausdorf dimensions equal
suggests that avalanches are {\it uncorrelated}.

\begin{figure}
\epsfxsize=7.5cm
\hskip0.5cm
\epsfbox{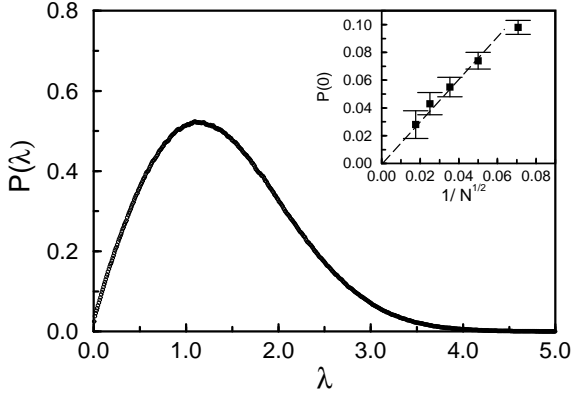}
\vskip0.01truecm
\caption{\label{fig:lambda}  
The distribution of the local stabilities, $P(\lambda)$, for $N = 3200$. Inset:
The finite size scaling of $P(0)$.}
\end{figure}

These results, in particular the size $N$ as the sole cutoff of the different
distribution functions, which all exhibit power law behaviour, confirm the 
above - stated {\it self-organized criticality} of the entire hysteresis 
loop of the SK model.
To shed more light on the underlying physics we explore the local-fields, 
$h_i=\sum_j J_{ij}\sigma_j+h$ by studying the local stabilities, 
$\lambda_i=\sigma_i h_i$, which are all positive for stable spin 
configurations. Their distribution, $P(\lambda)$ is shown in 
Fig.~\ref{fig:lambda}.
Remarkably -- unlike the local field distribution~\cite{bertotti} -- 
$P(\lambda)$ is essentially the same at any point of the 
hysteresis loop. This suggests that the avalanche dynamics of the SK model 
organizes the system into special states with similar properties
{\it everywhere} along the hystersis loop. A careful finite size analysis 
shows that $P(\lambda = 0) \sim 1/\sqrt{N}$ and $P(\lambda) \approx C 
\lambda^\alpha $ with $C\approx \alpha \approx 1$ for small 
$\lambda$'s. As we now show, this latter result establishes once again
that these special states are {\it critical}.
To prove this let us flip $n_{flip}$ arbitrary spins starting from a given 
stable spin configuration $\{\sigma_i\}$  with $\lambda_i > 0$  
and calculate the average number of new unstable spins,
$\langle n_{unst} \rangle $,
distinguished by negative stabilities $\lambda_i' = \lambda_i + 
\Delta\lambda_i < 0$: 
\begin{equation}
\lambda_i'=\lambda_i -2\sum_{j\;{\rm flipped}}\sigma_i J_{ij}\sigma_j.
\label{lambda}
\end{equation}  
The system is critical if $\langle n_{unst} \rangle = n_{flip}$,
as for $\langle n_{unst} \rangle < n_{flip}$ the avalanches die out 
exponentially fast while in the opposite case they explode~\cite{sethna}.
Assuming that the $n_{flip}$ random terms at the rhs. of Eq.~(\ref{lambda}) 
are independent, the probability $P_d$ of destabilizing a 
given spin is:
\begin{equation}
P_d = \int_0^\infty d\lambda\; P(\lambda) \int_{-\infty}^{-\lambda}
d(\Delta\lambda) \; Q(\Delta\lambda)  \;,
\end{equation} 
where $Q(\Delta\lambda) =  \exp\{ -N \Delta \lambda^2 /8 n_{flip}\} 
\sqrt{N/8\pi n_{flip}}$ is the probability distribution of the 
$\Delta\lambda$ term in 
Eq.~(\ref{lambda}), and $P(\lambda)$ is approximated by its asymptotic form, 
$P(\lambda) = C \lambda^\alpha$. The average number of destabilized spins is 
then $\langle n_{unst} \rangle = N P_d = \tilde C(\alpha) 
 N (n_{flip}/N) ^{(\alpha + 1)/2}$, with $\tilde C(\alpha)$ an 
$\alpha$-dependent 
constant, $\tilde C(1) = C$. For $\alpha > 1$ (or $\alpha = 1$ and $C< 1$) 
$\langle n_{unst} \rangle < n_{flip}$ and the system cannot give rise to large 
avalanches. On the other hand, for $\alpha < 1$ (or $\alpha = 1$ and $C> 
1$) $\langle n_{unst} \rangle > n_{flip}$, and the state is
unstable. Thus the criticality condition is characterized by $\alpha = 1$ 
and $C = 1$. These are exactly the values found in our numerical simulations,
once again underlining the criticality of the system.

The physical mechanism of self-organized criticality 
can be qualitatively understood as follows. As the avalanche rolls,
at any given time step $t$ the stabilities of the spins are shifted {\it only}
by those spins, which changed sign at step $t-1$. These spins have flipped
because the second term of Eq.~(\ref{lambda}) for their stabilities 
was negative, pulling their $\lambda_{i}$'s downward. However once 
$\lambda_{i}$ changed sign, the very same term now enhances this 
stability. More importantly, this term being symmetric, it also pushes upward 
the stabilities of the other spins of the avalanche, which pulled spin $i$
down and flipped it in the first place. This effect is suppressing
the density of states with low local fields, reminescent of the
formation of the Coulomb gap in the disordered electron problem~\cite{efros}.
The stabilities of the spins which did not participate in the avalanche
will be shifted by a random amount by the just-flipped spins. However in 
the presence of a slope in their distribution $P(\lambda)$, this will have a 
net effect, moving the stabilities of more spins downward than upward.
In short, {\it correlations} between the spins of an avalanche move
the stabilities of the already flipped spins {\it upward}; at the same time
the {\it random couplings} between all spins drive a net {\it downward} 
drift. The competition of these two forces keeps the system {\it critical}.

To understand the shape of the measured major hysteresis loop more in detail
we first observe that the states where an avalanche stops must 
always be {\em single spin-flip stable} (SSS). Let us therefore define the 
average number of SSS states, 
\begin{equation}
\langle V(m,h) \rangle = \left\langle {\rm Tr} \bigl\{
\prod_{i = 1} ^ N \Theta(\lambda_i) \;\delta(mN -\sum_i \sigma_i)\bigr\} 
\right\rangle\;,
\label{eq:V}
\end{equation}
where the angular bracket indicates annealed disorder averaging and the 
trace stands for the  summation over all  spin configurations. 
The product of the theta functions in Eq.~(\ref{eq:V})
projects out those states where all the spins have positive stabilities,  
while the delta function selects states with a given magnetization.
 
Using the integral representations $\delta(y) = 
\int_{-\infty}^\infty {dx\over 2\pi} e^{-ixy} $ and 
$\Theta(\lambda_i) = \int_{-\infty}^\infty {i dz_i\over 2\pi z_i} e^{-i z_i 
\lambda_i}$ the 
function $V(m,h)$ can be rewritten in an exponential form, 
$V(m,h) = \int {dx\over 2\pi} \prod_i \bigl[\int {i dz_i\over 2\pi z_i} 
\exp\{- i (\lambda_i z_i - x \sigma_i +mx) \} \bigr]$, 
and  the disorder average and the spin summation can be easily carried out. 
After the disorder  averaging the effective action contains a term 
proportional  to $\sim (\sum_i z_i)^2$.    Decoupling  this term with a new 
Hubbard-Stratonovich field $R$,  one finally  arrives at the following 
expression:
\begin{eqnarray}
&&\langle V \rangle  =  \sqrt{N} \int {dx\over 2\pi} 
\int {dR \over \sqrt{2\pi}} \;\exp\bigl\{ N [\ln Q  -{R^2 \over 2} - im x 
]\bigr\}\;, 
\nonumber \\
&&~~ Q(h,x, p)  =  \int {i dz\over 2\pi z} \;2\; \cos(x-zh) \;
e^{-{1\over2}z^2 -i R z} \; .
\label{eq:Q}
\end{eqnarray}
The above integral can readily be evaluated in the $N\to\infty$ limit 
by the saddle point method. The saddle point equation, $\partial Q 
/\partial x  = i m Q $ can be solved analytically,  and  the variable $x$ 
can be completely eliminated resulting in the following 
expression for   $\langle V(m,h)\rangle$:
\begin{eqnarray}
&&\phantom{nnn} \langle V(h,m) \rangle \sim \exp \bigl( N \Omega_{sp}(m,h)
\bigr)\;, \nonumber\\
&& \Omega_{sp} = {1-m\over 2} \ln{1 - \phi^- \over 1-m} + {1+m\over 2} 
\ln{1 + \phi^+ \over 1+m} - {R^2\over 2} \;, \nonumber 
\end{eqnarray}
where $R$ is determined from
$\partial~ \Omega_{sp}(m,h,R)/\partial~ R = 0$,
and $\phi^{\pm} = \phi({h\pm R \over \sqrt{2}})$ with 
$\phi(x) = {2\over \sqrt{\pi} }  \int_0^x e^{-t^2} dt$. 
\begin{figure} 
\hskip 0.5cm
{\psfig{figure=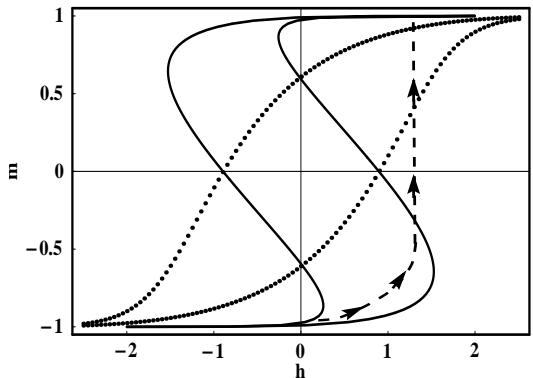,height=5.0cm,width=7.0cm,angle=00}}
\caption{Outer bound of the region of the single spin-flip stable states. 
Dotted line: $J_0=0$; solid line: $J_0=2.5$. The impossiblity of a monotonic 
$m(h)$ curve within these bounds forces the jump, as indicated by the arrows.}
\end{figure}

In Fig.4. we plotted the contour of $\Omega_{sp} = 0$. 
Outside this line the density of SSS states scales to $0$ exponentially, 
thus they are definitely unable to arrest the avalanches.
Inside this line the number of SSS states is exponentially large, and is 
thus comparable to the total number of states, themselves 
expoential in $N$. Therefore avalanches get trapped with a higher 
probability in one of the SSS states. 
Hence the $\Omega_{sp} = 0$ contour constitutes a strict outer bound for the 
true hysteresis loop.

Comparing Figs.1 and 4 shows that the $\Omega_{sp} = 0$ contour considerably
overestimates the size of the hysteresis loop.
We pursued two refinements of this calculation. We developed a replica 
symmetric description as well as a de Almeida - Thouless type replicon
instability analysis: these will be reported separately~\cite{next}.

Finally, we briefly discuss the effect of a finite ferromagnetic coupling, 
$J_0>0$. $J_0$ simply shifts the value of the magnetic 
field  $h \to h + J_0 m$ in Eq.~(\ref{eq:Q}), and results 
in a {\em shear } of the entire contour $\Omega_{sp} = 0$. For a branch of the 
hysteresis loop $m$ must be a monotonic function of 
$h$. Since the $\Omega_{sp} = 0$ contour is an outer bound of the 
hysteresis loop, therefore, when this loop would force a non-monotonic $m(h)$ 
relation (Fig.4), the major hysteresis loop must develop a {\em finite jump}. 
Since the slope of the major hysteresis 
loop for $J_0 =0$ is finite, one expects this transition to
occur at a finite critical coupling, $J_0 = J_c$. Our numerical data 
agree with this picture~\cite{next}.

We end with a comparison to the random field Ising model.
In that model our initial simple scaling considerations
yield $\langle S \rangle \propto {\cal O}(1)$, i.e.
a non-critical avalanche distribution.
As shown in Ref.\cite{sethna}, there is only a single critical point
at the coercive field at some specific value of the disorder.
Therefore the distribution functions exhibit a scaling behaviour, with
the cutoff set by the distance from this critical point, rather
than by the system size, as happens for the SK model.
In short, the RFIM exhibits ``plain old criticality''~\cite{sethna},
while the SK model exhibits self-organized criticality.
Also, the avalanche distribution exponents $\tau$ in the two models are 
different: on the mean field level $\tau=1.5$ for the RFIM and $1.0$ 
for the SK model. In finite dimensions the exponents typically increase:
numerical simulations of the 3D RFIM found
$\tau = 1.6$~\cite{sethna}. In contrast, numerical studies of realistic
3D models~\cite{num}, as well as experimental~\cite{expt} works
report $\tau$ in the $1.1 ... 1.4$ regime. This raises the possibility
that the finite dimensional extensions of the frustrated models might provide
a $\tau$ closer to the experimental values.

In sum, we studied the hysteretic behaviour of the SK model.
We determined numerically the distribution functions of the avalanches,
the magnetization jumps, and the local fields. The model exhibits 
{\it self-organized criticality everywhere along the hysteresis loop}.
We recalculated the loop with analytic methods
as the location of one-spin-flip stable states, and found satisfactory
agreement with the numerical results.

We acknowledge useful discussions with R. Scalettar, G. Bertotti, E. Della 
Torre, G. K\'ad\'ar, M. M\'ezard and J. Sethna. This research has been 
supported by the U.S - Hungarian Joint Fund 587, Hungarian Grants OTKA~T026327 
and OTKA~D29236 and by NSF DMR 95-28535. GTZ acknowledges the kind hospitality 
of the George Washington University.

\end{document}